\documentstyle[aps,fleqn,floats]{revtex}

\begin{document}
%\draft
\twocolumn[\hsize\textwidth\columnwidth\hsize\csname @twocolumnfalse\endcsname

\title{Excitons with anisotropic effective mass}
\author{Arno Schindlmayr\cite{email}}
\address{Cavendish Laboratory, University of Cambridge, Madingley Road,
Cambridge CB3 0HE, UK}
\date{Received 25 February 1997, in final form 19 May 1997}
\maketitle

\begin{abstract}
{\bf Abstract.}
We present a simple analytic scheme for calculating the binding energy of
excitons in semiconductors that takes full account of the existing
anisotropy in the effective mass, as a complement to the qualitative
treatment in most textbooks. Results obtained for excitons in gallium
nitride form the basis for a discussion of the accuracy of this approach.
\end{abstract}

\begin{abstract}
{\bf Zusammenfassung.}
Wir pr\"asentieren ein einfaches analytisches Verfahren zur Berechnung der
Bindungsenergie von Exzitonen in Halbleitern, das die vorhandene
Anisotropie in der effektiven Masse vollst\"andig miteinbezieht, in
Erg\"anzung zu der qualitativen Betrachtung in den meisten Lehrb\"uchern.
Ergebnisse f\"ur Exzitonen in Galliumnitrid bilden die Grundlage f\"ur eine
Diskussion der Genauigkeit dieser Methode.
\end{abstract}

]
\narrowtext

\section{Introduction}

The traditional one-particle theory of semiconductors as taught in
undergraduate courses is that of a material with a finite energy gap
separating the highest occupied from the lowest unoccupied electronic
state. In this picture the minimum energy for an elementary excitation is
that required to raise a valence band electron into the conduction band,
and is thus equal to the gap $E_{\rm g}$. However, the electron and the
hole, which is created in the same process, need not separate completely
and can instead form a bound pair under the influence of their mutual
Coulomb attraction \cite{Wan37}. Such bound electron-hole pairs, which
transport energy and momentum but no charge, are called {\em excitons}.
They are, in fact, the truly lowest elementary excitations of a pure
semiconductor and their influence on the optical properties of a material
is profound. Most importantly, the occurrence of excitons lowers the
threshold for photon absorption to $E_{\rm g} - E_{\rm b}$, where
$E_{\rm b}$ denotes the internal binding energy of the electron-hole pair.

Because of their practical significance excitons feature in all textbooks
on solid state physics, but the discussion is usually restricted to some
qualitative arguments based on formal similarities to the hydrogen atom
problem, which is modified only by a dielectric constant to account for the
surrounding medium and by empirical effective masses for the electron and
hole \cite{Ash76}. The latter are always assumed to be isotropic, but while
there certainly are textbook examples such as CdS for which this condition
is nearly satisfied and hydrogenic absorption series have indeed been
observed \cite{Hop61}, in most semiconductors the anisotropy in the
effective mass is so large that it cannot be ignored in a quantitative
treatment. In this paper we present a variational scheme for calculating
the binding energy $E_{\rm b}$ of excitons in realistic materials that
takes full account of the existing anisotropy. Our priority has been to
maintain a universally applicable and strictly analytic approach suitable
for teaching purposes.

\section{Variational exciton wavefunction}

Most semiconductors used for modern electronic devices crystallize in the
diamond (e.g.\ Si, Ge), zincblende (e.g.\ GaAs) or wurtzite (e.g.\ GaN)
structure, for which the constant energy surfaces $E({\bf k})$ about the
valence band maximum and conduction band minimum are ellipsoidal in shape,
yielding distinct longitudinal electron and hole effective masses
$m^{\rm e}_\parallel$ and $m^{\rm h}_\parallel$ along one principal axis,
and transverse effective masses $m^{\rm e}_\perp$ and $m^{\rm h}_\perp$ in
the plane perpendicular to it. Here we will focus on this geometry,
although the method is readily generalized. Taking the principal axis in
the $z$-direction, the Hamiltonian for the relative motion of the
electron-hole pair is
\begin{eqnarray} \label{hamiltonian}
H &=& -\frac{\hbar^2}{2 \mu_\perp} \left( \frac{\partial^2}{\partial x^2}
+ \frac{\partial^2}{\partial y^2} \right) -\frac{\hbar^2}{2 \mu_\parallel}
\frac{\partial^2}{\partial z^2} \nonumber \\
&&- \frac{e^2}{4 \pi \epsilon_0 \epsilon \sqrt{x^2 + y^2 + z^2}}
\end{eqnarray}
where $\mu_\perp = m^{\rm e}_\perp m^{\rm h}_\perp / (m^{\rm e}_\perp +
m^{\rm h}_\perp)$ and $\mu_\parallel = m^{\rm e}_\parallel
m^{\rm h}_\parallel / (m^{\rm e}_\parallel + m^{\rm h}_\parallel)$ denote
the reduced transverse and longitudinal effective mass, respectively,
and $\epsilon$ is a suitable dielectric constant. The anisotropy destroys
the spherical symmetry of the hydrogen Hamiltonian, yielding a wavefunction
with a different characteristic localization along the principal axis and
in the transverse plane. We therefore choose a generalization of the
hydrogen ground-state wavefunction with ellipsoidal symmetry
\begin{equation} \label{wavefunction}
\hspace{-\mathindent} \psi(x,y,z) = \left( \frac{\beta^3}{\lambda \pi}
\right)^{1/2} \exp \left(-\beta \sqrt{x^2 + y^2 + (z/\lambda)^2} \right)
\end{equation}
where the two parameters $\beta$ and $\lambda$ can be varied independently
to control the transverse and longitudinal extension. The wavefunction, of
course, becomes exact in the isotropic case $\mu_\parallel = \mu_\perp$
with $\lambda = 1$ and $\beta^{-1} = a_0 \epsilon m / \mu_\perp$, where
$a_0 = 0.529\,{\rm \AA}$ denotes the Bohr radius and $m$ is the free
electron mass, but it remains an excellent approximation even when the
anisotropy is large. A wavefunction of the type (\ref{wavefunction}) was
originally proposed in \cite{Kit54} to describe shallow donor impurity
states in Si and Ge, but treated numerically and evaluated only for the
mathematically distinct case $\lambda < 1$, reflecting the fact that the
longitudinal effective mass is greater than the transverse one for
electrons at the bottom of the conduction band in both materials. No such
restriction will be made in this paper.

Given the explicit form of the wavefunction (\ref{wavefunction}) the
calculation of the kinetic energy is straightforward, using the
substitution $z' = z / \lambda$ and a subsequent transformation to
spherical coordinates. We obtain
\begin{eqnarray}
-\frac{\hbar^2}{2 \mu_\perp} \int\!\! \psi \frac{\partial^2}{\partial x^2}
\psi \,{\rm d}^3r &=& -\frac{\hbar^2}{2 \mu_\perp} \int\!\! \psi
\frac{\partial^2}{\partial y^2} \psi \,{\rm d}^3r \nonumber \\
&=& \frac{\hbar^2}{6 \mu_\perp} \beta^2
\end{eqnarray}
for the contribution in the transverse isotropic plane, and similarly
\begin{equation}
-\frac{\hbar^2}{2 \mu_\parallel} \int\!\! \psi \frac{\partial^2}{\partial
z^2} \psi \,{\rm d}^3r = \frac{\hbar^2}{6 \lambda^2 \mu_\parallel} \beta^2
\end{equation}
for the relative motion along the principal axis. To calculate the
potential energy we use the same transformation to spherical coordinates.
The integrals over the radial variable and the azimuth angle are readily
evaluated, leaving
\begin{eqnarray}
\lefteqn{ -\frac{e^2}{4 \pi \epsilon_0 \epsilon} \int\!\! \psi
\frac{1}{\sqrt{x^2+y^2+z^2}} \psi \,{\rm d}^3r } \nonumber \\
&&= -\frac{e^2}{8 \pi \epsilon_0 \epsilon} \beta \int_0^\pi\!
\frac{\sin\theta \,{\rm d}\theta}{\sqrt{1 + (\lambda^2 - 1) \cos^2\theta}}
\;.
\end{eqnarray}
The evaluation of the remaining integral over the polar angle depends on
the sign of the factor $\lambda^2 - 1$. If $\lambda > 1$ we substitute
$t = \sqrt{\lambda^2 - 1} \cos\theta$, otherwise we use the substitution
$t = \sqrt{1 - \lambda^2} \cos\theta$ to obtain an elementary integral,
which is solved by
\begin{eqnarray}
I(\lambda) &=& \frac{1}{2} \int_0^\pi\! \frac{\sin\theta \,{\rm d}\theta}
{\sqrt{1 + (\lambda^2 - 1) \cos^2\theta}} \nonumber \\
&=& \left\{
\begin{array}{lcl}
\displaystyle \frac{\mbox{arcsinh}\,\sqrt{\lambda^2-1}}{\sqrt{\lambda^2-1}}
& \mbox{for} & \lambda > 1 \\[2ex]
\displaystyle \frac{\arcsin\sqrt{1-\lambda^2}}{\sqrt{1-\lambda^2}}
& \mbox{for} & \lambda < 1.
\end{array}
\right.
\end{eqnarray}
While we have to make this formal case distinction, we emphasize that the
energy function is smooth, with both branches of $I(\lambda)$ approaching
unity in the limit $\lambda \to 1$.

Collecting the kinetic and potential contributions, we thus obtain the
expression
\begin{equation} \label{energyfun}
E(\beta,\lambda) = \frac{\hbar^2}{6} \beta^2
\left( \frac{2}{\mu_\perp} + \frac{1}{\lambda^2 \mu_\parallel} \right)
- \frac{e^2}{4 \pi \epsilon_0 \epsilon} \beta I(\lambda)
\end{equation}
for the ground state energy, which must be minimized with respect to the
parameters $\beta$ and $\lambda$. The kinetic term is quadratic in $\beta$
while the potential term is linear, so the respective partial derivative is
readily performed. The condition $\partial E / \partial \beta = 0$ then
yields a relation between the two parameters at the energy minimum
\begin{equation} \label{beta}
\beta = 3 \left( \frac{e^2}{4 \pi \epsilon_0 \epsilon \hbar} \right)^2
\left( \frac{2}{\mu_\perp} + \frac{1}{\lambda^2 \mu_\parallel} \right)^{-1}
I(\lambda)
\end{equation}
which, when substituted in (\ref{energyfun}), allows us to rewrite the
energy as a function of $\lambda$ only
\begin{equation} \label{energy}
\hspace{-\mathindent} E(\lambda) = -\frac{3}{2} \left( \frac{e^2}{4 \pi
\epsilon_0 \epsilon \hbar} \right)^2 \left( \frac{2}{\mu_\perp} +
\frac{1}{\lambda^2 \mu_\parallel} \right)^{-1} I(\lambda)^2.
\end{equation}
The energy minimum is found at the stationary point for which $\partial E /
\partial \lambda = 0$ and through simple mathematical rearrangement we can
express this condition in the form
\begin{equation} \label{graph}
\frac{\mu_\perp}{\mu_\parallel} = 2 \lambda^3
\frac{1 - \lambda I(\lambda)}{I(\lambda) - \lambda}
\end{equation}
which may be solved graphically. The important point to note is that the
right-hand side of (\ref{graph}) is a universal function $f(\lambda)$ that
does not depend on the material properties, so the same plot can be used
for all semiconductors to determine $\lambda$. In practice, however, the
reduced transverse and longitudinal effective mass will often not differ by
more than a factor of three and in this value range the function on the
right-hand side of (\ref{graph}) is accurately approximated by its
lowest-order polynomial term $f(\lambda) \approx \lambda^3$. The parameter
$\lambda$ is then explicitly given by
\begin{equation} \label{lambda}
\lambda = \left( \frac{\mu_\perp}{\mu_\parallel} \right)^{1/3}.
\end{equation}
This simplification allows for a very efficient analytic calculation of the
material-specific binding energy $E_{\rm b}$, which is given by the modulus
of the ground state energy according to (\ref{energy}). It is still exact
in the isotropic limit $\mu_\parallel = \mu_\perp$, yielding the correct
binding energy $E_{\rm b} = R_\infty m / (\mu_\perp \epsilon^2)$ where
$R_\infty = 13.6\,{\rm eV}$ is the hydrogenic Rydberg energy. The
wavefunction itself, required for instance to calculate the optical
absorption coefficient, is given by the original variational expression
with $\beta$ defined through the relation (\ref{beta}).

\section{Numerical results for G\lowercase{a}N}

In order to illustrate the numerical quality of our scheme, we now consider
the case of gallium nitride as an explicit example for excitonic binding
energies in a realistic semiconductor. In GaN the valence band maximum
comprises three almost degenerate subbands, giving rise to three distinct
hole types that can partake in the formation of excitons. Conventionally,
these are referred to as light holes, heavy holes and split-off holes,
reflecting their different effective masses. Within the same material we
can study excitons for which the reduced longitudinal effective mass is
greater than the transverse as well as excitons for which it is smaller,
and test the accuracy of the approximate treatment in either case.

\begin{table}
\caption{Comparison between approximate and numerically calculated binding
energies $E_{\rm b}$ for excitons in GaN, formed with holes of different
effective mass $m^{\rm h}$ from the three nearly degenerate subbands at the
valence band maximum. The analytic approximation is accurate even when the
anisotropy $\mu_\perp/\mu_\parallel$ is very different from unity.}
\label{tab:results}
\begin{tabular}{lccccc}
& & & & \multicolumn{2}{c}{$E_{\rm b}$ (meV)} \\
\cline{5-6}
Hole type & $m^{\rm h}_\perp$ & $m^{\rm h}_\parallel$ & $\mu_\perp /
\mu_\parallel$ & approx. & numer. \\
\hline
Light hole     & 0.15 & 1.10 & 0.483 & 15.456 & 15.458 \\
Heavy hole     & 1.65 & 1.10 & 0.959 & 24.809 & 24.809 \\
Split-off hole & 1.10 & 0.15 & 1.805 & 18.939 & 18.941 \\
\end{tabular}
\end{table}

The effective mass of the electron at the bottom of the conduction band is
$m^{\rm e}_\perp = 0.18$ in the transverse and $m^{\rm e}_\parallel = 0.20$
in the longitudinal direction, given in units of the free electron mass
$m$, while corresponding parameters for the three hole types are listed in
table \ref{tab:results}. All values are quoted from \cite{Suz95}. Next the
calculated anisotropy factors $\mu_\perp / \mu_\parallel$ are given. For
excitons formed with heavy holes this is close to unity, indicating a
rather small perturbation from the isotropic case, but the ratio is
substantially smaller with light holes and larger with split-off holes,
respectively. The fifth column in table \ref{tab:results} lists the binding
energies $E_{\rm b}$ obtained analytically from (\ref{energy}) with
$\lambda = ( \mu_\perp / \mu_\parallel )^{1/3}$. These are compared with
results that we obtained by exact diagonalization of the Hamiltonian
(\ref{hamiltonian}) using standard numerical techniques. The applicable
value of the static dielectric constant is the low-frequency limit
$\epsilon = 9.5$ of GaN.

The excellent agreement between the approximate and numerical results
in all cases confirms the validity of the variational wavefunction
(\ref{wavefunction}) as well as the accuracy of the additional
simplification (\ref{lambda}) that makes the approach strictly analytic.
Comparison with experimental data is more problematic because it is very
difficult to extract accurate binding energies from optical spectra.
Nevertheless, the recently published values of 21 meV for excitons with
light and heavy holes and 23 meV for excitons with split-off holes in GaN
are probably quite reliable \cite{Sha96}. The discrepancy with the results
obtained here is due to the underlying model Hamiltonian
(\ref{hamiltonian}), however, not the analytic approximations introduced to
solve it. While the anisotropy in the effective mass is treated adequately,
other important features such as the mixing of states at the threefold
degenerate valence band maximum or the spatial variation of the dielectric
function are still neglected. A more involved quantitative scheme will also
have to incorporate these in order to reproduce experimental binding
energies for real materials.

\section*{Acknowledgements}

The author wishes to thank T Uenoyama and M Suzuki for inspiration and
hospitality at the Central Research Laboratories of Matsushita Electric
Industrial Co., Ltd.\ in the summer of 1996, and R W Godby for useful
discussions. Financial support from the Deutscher Akademischer
Austauschdienst under its HSP III scheme, the Studienstiftung des deutschen
Volkes, the Gottlieb Daimler- und Karl Benz-Stiftung, Pembroke College
Cambridge, the Engineering and Physical Sciences Research Council, and the
Japan International Science and Technology Exchange Center is gratefully
acknowledged.

\end{document}